\documentclass[a4paper,twoside]{article}

\usepackage{epsfig}
\usepackage{subfigure}
\usepackage{calc}
\usepackage{amssymb}
\usepackage{amstext}
\usepackage{amsmath}
\usepackage{multicol}
\usepackage{pslatex}
\usepackage{apalike}
\usepackage{fancyhdr}
\usepackage{insticc}
\usepackage[small]{caption}

\begin{document}
\onecolumn \maketitle \normalsize \vfill

\section{\uppercase{Introduction}}
\label{sec:introduction}

\noindent Blind source separation (BSS) is an active research topic of signal and image processing in recent years. It considers separating a set of unknown signals from their observed mixtures, with reasonable assumptions of the form of the mixing process: linear or nonlinear, instantaneous or convoluting, under or over determined, noisy or noiseless, and so on. However, in all cases the mixing coefficients remain unknown and have to be estimated as well as original source signals.

Various methods and models have been proposed for BSS task, among which Principal Component Analysis (PCA) seeks orthogonal directions of maximum variance exhibited by the data as source axes, while Independent Component Analysis (ICA) \cite{ICA2001}, in its basic form, assumes statistical independency of sources and linear mixing process and consists of seeking an inverse linear transformation matrix applying on the data to achieve maximum mutual independency between output components. Both methods exploits basic statistical characteristics of source signals to achieve the separation, which makes them well generalizable and robust in cases that as few prerequisite assumptions as uncorrelatedness or independency can be made about the source. Some variant algorithms are also proposed to adapt to certain relaxation of model assumptions like nonlinearity or noises \cite{Harmeling_NC03,Almeida_JMLR05}. However, in many other cases, we may find the availability or the needs of various types of prior information to regulate the essentially ill-posed BSS problem. Compared with PCA and ICA, Bayesian framework allows convenient introduction of these prior constraints about the sources and the mixing coefficients, and more important, supports flexible structuring and integrating multiple hierarchical clues for separation purpose.

In the field of image processing, BSS approaches are being widely employed to separate or segment mixed images observed from, for example, satelite and hyper-spectral imaging \cite{Hichem_JEI04,Parra_NIPS00,Miguel_IWANN03}, medical imaging \cite{Calhoun_EMBM98,Snoussi_ICIP05}, and other superimpositions of natural images \cite{Brons_JMLR05,Castella_LNCS04}.

\begin{figure}[!h]
  \centering
  \includegraphics[scale=0.5]{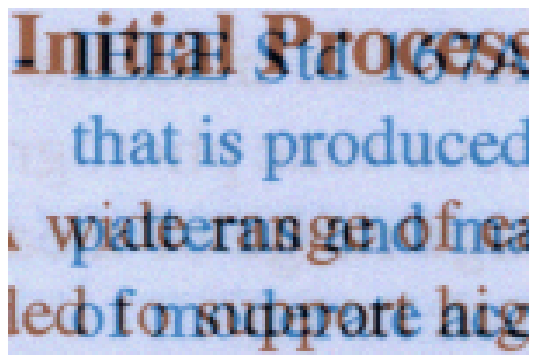} 
  \includegraphics[scale=0.45]{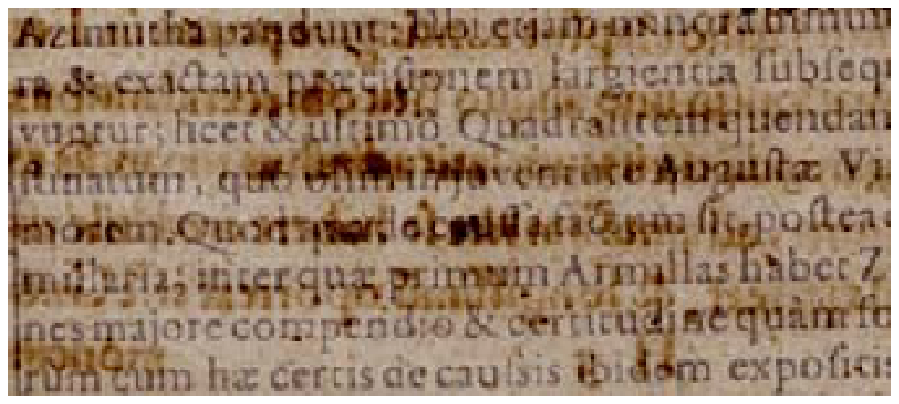}
  \\
  (a) \hspace{3cm} (b)
  \caption{Examples of mixed document images: a) show-through mixture; b) bleed-through mixture.}
  \label{fig:mixsamp}
\end{figure}

This paper focuses on one specific type of images - document images, where superimposition of two images usually appears as a major type of degradation encountered in digitization \cite{Sharma_TIP01} or ancient documents \cite{Drira_DIAL06}. The former degradation usually occurs as artifact during scanning a double-sided  document when the text on the back-side printing shows through the non-opaque medium and are mixed with the foreside text. Fig.\ref{fig:mixsamp}a shows one such example. The latter cause of text superimposition, usually called bleed-through, can often be observed in old documentations due to ink blurring or penetrating as illustrated by Fig.\ref{fig:mixsamp}b. Other forms of overlapped patterns, like underwriting and watermarks, are also common. Though the actual underlying mixing process may be quite complicated and diverse in various mixture forms, the linear mixing model usually serves as a resonable approximation and benefits analytical and computational simplicity, thus is adopted in most document separation cases. 

To separate document image mixtures, the common PCA and ICA algorithms can be used and have shown their effectiveness in detecting independent document features like watermarks, as inspected in \cite{Anna_ICIAR04} where each source was considered as random signal sequence in a whole without further internal structuring. The Bayesian framework has also been used before for document separation as in \cite{Anna_TIP06}, where the source is modeled by a Markov Random Field on the pixel values to account for local smoothness inside one object,  as well as an extra line process enforcing the discontinuity at object edges.

In this contribution, we propose a solution to jointly separate and segment linearly mixed document images. Besides considering the mixture in single grayscale channel, we address the joint separation of multi-channel mixture of multiple sources. In section 2, we give the probability formulation of the problem. In section 3, the algorithm of Bayesian estimation for model parameters is described. In section 4, simulation results of the proposed algorithm are shown on both synthetic and real images.

\section{\uppercase{Model Assumption and Formulation}}

\noindent Document images are created by various digitization methods from vast types of documentation. Commonly, a color scanner can be used to produce three different views of one document in the red, green, and blue channels. With detectors working in non-visible wavelengths such as infrared and ultraviolet, even more information channels of data can be obtained, depending on the object of interest in documents.

Given observations of $M$ different mixtures, either in grayscale or multiple channels, our work is thus to obtain $N$ corresponding source images (normally $M>N$) in the same pixel format as the observations.

\subsection{Data Model}

In this work, the observations are $M$ registered images $(X_i)_{i=1 \ldots M}$, which are defined on the same set of pixels $\mathcal{R}$: $X_i=\{x_i(r)\}_{r\in\mathcal{R}}$. The observations are noisy linear instantaneous mixture of $N$ source images $(S_j)_{j=1 \ldots N}$ also defined on $\mathcal{R}$, following the data generation model given by:
\begin{equation}\label{eqn:mix}
\mathbf{x}(r)=\mathbf{A}\mathbf{s}(r)+\mathbf{n}(r) \qquad r\in\mathcal{R}
\end{equation}
where $A=(a_{ij})_{M \times N}$ is the unknown mixing matrix, $\mathbf{n}(r)$ is a set of independent zero-mean white Gaussian noise for each observation with variance $\mathbf{\sigma}^2_{\epsilon} = (\sigma^2_{\epsilon 1}\ldots\sigma^2_{\epsilon M})$, $\mathbf{x}(r)$ and $\mathbf{s}(r)$ are the observation and source vector at pixel $r$ respectively. Let $\mathbf{S}=\{\mathbf{s}(r),r\in\mathcal{R}\}$, $\mathbf{X}=\{\mathbf{x}(r),r\in\mathcal{R}\}$, and denote the noise covariance matrix by $\mathbf{R}_\epsilon=diag[\sigma^2_{\epsilon 1}\ldots\sigma^2_{\epsilon M}]$, 
we have the Gaussian distribution for the observations given the sources and the mixing parameters:
\begin{equation}\label{eqn:datagaus}
p(\mathbf{X}|\mathbf{S},\mathbf{A},\mathbf{R}_\epsilon) = \prod_r \mathcal{N}(\mathbf{A}\mathbf{s}(r),\mathbf{R}_\epsilon)
\end{equation}

\subsection{Source Model}

We model the distribution of pixel intensity for each source images (and for each color channel) by a Mixture of Gaussians (MoG), whose components correspond to each object type (or class) that appears roughly equal pixel values. For example, the simplest model may consist of two components, one for foreground text and the other for background blank. Furthermore, to allow imposing constraints on distribution of class labels, for every source $S_j$ we represent the class labels by a set of discrete hidden variables $Z_j=\{z_j(r),r \in \mathcal{R}\}$ with $z_j(r)$ taking values from $\{1,\ldots,K_j\}$, where $K_j$ is the total number of classes in image $S_j$. In the following, we assume all $\{K_j\}$ equal to the same value $K$.

Given pixel labels, pixels of different classes can be reasonably assumed independent, while concerning the pixels inside a given class, there are usually two choices:
\begin{enumerate}
\item We may assume pixel intensities are conditionally independent given their labels; 
\item Alternatively, we may explicitly take into account the local dependency between neighboring pixels of same class.
\end{enumerate}

In the first choice, the distribution of pixel $r$ in the $j^{th}$ source is modeled by: 
\begin{equation}\label{eqn:srclab}
p(s_j(r)|z_j(r)=k)=\mathcal{N}(\mu_{jk},\sigma^2_{jk})
\end{equation}
where $\mu_{jk}$ and $\sigma^2_{jk}$ are the mean and variance of the $k^{th}$ Gaussian component of the $j^{th}$ source. Assuming independency between different sources and denoting the set of labels corresponding to every source by $\mathbf{Z}=\{Z_j,j=1 \ldots N\}$, we have:
$$ p(\mathbf{S}|\mathbf{Z}) = \prod_j\prod_k\prod_{\{r:z_j(r)=k\}} p(s_j(r)|z_j(r)=k) $$
which is by (\ref{eqn:srclab}) also a Gaussian and spatially separable on $r$.

In the second choice, the local dependency can be accounted by extra smoothness constraints, like the mean value, between neighboring pixels. We first assign a binary valued contour flag $q_j(r)$ for every pixel $r$ of every source $j$, which is deterministicly computed by:
$$ q_j(r) = \left\{ \begin{array}{l} 
1 \qquad \mbox{if} \quad z_j(r')=z_j(r), \forall r' \in \mathcal{V}(r) \\ 
0 \qquad \mbox{else} \\
\end{array} \right. $$ 
where $\mathcal{V}(r)$ denotes the neighbor sites of the site $r$.

Then, based on the value of the contour flag and possibly current values of the neighboring pixels, the distribution of intensity of individual pixel is formulated as:
\begin{equation}\label{eqn:srclabdep}
p(s_j(r)|z_j(r)=k,s_j(r'),r' \in \mathcal{V}(r)) = \mathcal{N}(\bar{s}_j(r),\bar{\sigma}^2_j(r))
\end{equation}
with,
\begin{align*}
\bar{s}_j(r) &= q_j(r)\mu_{jk} + (1-q_j(r))\frac{1}{|\mathcal{V}_{jk}(r)|} \sum_{r' \in \mathcal{V}_{jk}(r)}s_j(r') \\
\bar{\sigma}^2_j(r) &= q_j(r)\sigma^2_{jk} + (1-q_j(r))\sigma^2_j \\
\end{align*}
where $\mathcal{V}_{jk}(r)$ denotes the intersection of $\mathcal{V}(r)$ with the site set $\mathcal{R}_{jk}=\{r:z_j(r)=k\}$, $\sigma^2_j$ is the \emph{a prior} variance of pixel values inside a region. Eqn.(\ref{eqn:srclabdep}) states that at the contour pixel intensities follow the Gauss distribution whose parameters are determined by the class labels as (\ref{eqn:srclab}), while inside a region the distribution parameters are computed from the neighboring pixels. Note that under this assumption, $ p(\mathbf{S}|\mathbf{Z})$ is no longer separable on $r$, but with the \emph{parallel Gibbs sampling} scheme proposed in \cite{Feron_JEI05}, it can still be simulated efficiently.

As a commonly observed property of visual objects, pixels belonging to the same object usually connect to each other in a neighborhood of space, forming several connected regions of uniformly classified pixels, for instance, the multiple components constituting a text. By class labels defined earlier, this implies regional smoothness of the spatial distribution of class labels. This can be naturally modeled by \emph{a prior} Potts Markov Random Field for every label process $z_j(r)$:
\begin{equation}\label{eqn:potts}
p(z_j(r),r \in \mathcal{R}) \propto \exp \left[ \beta_j \sum_{r \in \mathcal{R}} \sum_{r' \in \mathcal{V}(r)} \delta(z_j(r)-z_j(r'))\right]
\end{equation}
The parameter $\beta$ reflects the degree of smoothing interactions between pixels and controls the expected size of the regions. In our work, all $\{\beta_j\}$ are assumed equal and assigned an empirical value within $[1.5,2.0]$.

\subsection{Multiple Channels}

When multi-channel image data are considered, there are multiple options for the processing model. We can perform separation of sources independently in each channel and by some measures merge the results in the end. Or, we may consider joint demixing for all channels. In the latter case, the mixing model can still have more alternatives:
\begin{enumerate}
\item[a)] all channels are equally mixed with the same mixing matrix; 
\item[b)] the mixing occurs separately in each channel with different mixing matrices; 
\item[c)] cross-channel mixing is assumed to be present.
\end{enumerate}
In the case of a), samples from different channels of the same observation can be concatenated for estimation of the mixing coefficients, which is similar to the monochrome case. In the case of c), an expanded mixing matrix $\mathbf{A}_{ML \times NL}$ (supposing $L$ channels) is used for all channels of all sources.

In this work, we assume the model b), where the mixing in different channels are mutually 
independent and with their own separate coefficients. Thus, in RGB color format, the sources and observations are actually $\{S_j^r,S_j^g,S_j^b\}_{j=1 \dots N}$ and $\{X_i^r,X_i^g,X_i^b\}_{i=1 \dots M}$. Correspondingly, there are $\{\mathbf{A}^r,\mathbf{R}_\epsilon^r,\mu_{jk}^r, \ldots \mathbf{A}^b,\mathbf{R}_\epsilon^b,\mu_{jk}^b\}$ and so on. But for each source $S_j$, only one classification field $Z_j$ is maintained and shared by all channels, as a natural way to enforce the common segmentation among different channels. This two-level hierarchical source model, which also facilitates introducing segmentation constraints like discontinuity and local regional dependency, is the main difference with the work of \cite{Anna_TIP06}, where a one-level MRF modeling of sources is defined on the single-channel pixel intensities along with an explicit binary edge process.

\section{\uppercase{Bayesian Estimation of Model Parameters}}

\noindent The unknown variables we want to estimate in the models given above are $\{\mathbf{S},\mathbf{Z},\mathbf{A},\mathbf{\theta}\}$, $\mathbf{\theta}$ representing all hyperparameters. The Bayesian estimation approach consists of deriving the posterior distribution of all the unknowns 
given the observation and then based on this distribution, employing appropriate estimators such as Maximum A Posteriori (MAP) or the Posterior Means (PM) for them. 
With our model assumptions, this posterior distribution can be expressed as: 
\begin{equation}\label{eqn:jointpdf}
p(\mathbf{S},\mathbf{Z},\mathbf{\theta}|\mathbf{X}) \propto 
p(\mathbf{X}|\mathbf{S},\mathbf{A},\mathbf{R}_\epsilon) 
p(\mathbf{S}|\mathbf{Z},\mathbf{\theta_s}) 
p(\mathbf{Z})p(\mathbf{\theta})
\end{equation}
where, $\mathbf{\theta_s}=\{(\mu_{jk},\sigma^2_{jk}),j=1 \ldots N,k=1 \ldots K\}$ and $\mathbf{\theta}=\{\mathbf{A},\mathbf{R}_\epsilon,\mathbf{\theta_s}\}$.

\subsection{Prior Assignments for Model Parameters}

According to the linear mixing model and all Gaussian assumptions, we choose corresponding conjugate priors for model hyperparameters. 
\begin{itemize}
\item Gaussian for source means $$\mu_{jk} \sim \mathcal{N}(\mu_{k0},\sigma^2_{k0})$$
\item Inverse Gamma for source variances $$\sigma^2_{jk} \sim \mathcal{IG}(\alpha_{k0},\beta_{k0})$$ 
\item Inverse Wishart for noise covariance $$\mathbf{R}_\epsilon^{-1} \sim \mathcal{W}_i(\alpha_{\epsilon_0},\beta_{\epsilon_0})$$
\end{itemize}
In this work, we assign uniform prior to $\mathbf{A}$ for simplicity and no preference of the mixing coefficients, while in other cases prior distributions like Gamma may be used to enforce positivity.

\subsection{Estimation by MCMC Sampling}

Given the joint \emph{a posteriori} distribution (\ref{eqn:jointpdf}) of all unknown variables, 
we use the Posterior Means as the estimation for them. Since direct integration over $\mathbf{z}$ is intractable, MCMC methods are employed in the actual Bayesian computations. In our work, a Gibbs sampling algorithm is used to generate a set of samples for every variable to be estimated, 
according to its full-conditional \emph{a posteriori} distribution given all other variables fixed to their current values. Then, after certain burn-in runs, sample means from further iterations are used as the Posterior Means estimation for the unknowns. The algorithm takes the form: 
\begin{itemize}
\item[ ]Repeat until converge,
\item[ ]
\begin{enumerate}
\item simulate $\mathbf{S}' \sim p(\mathbf{S}|\mathbf{Z},\mathbf{\theta},\mathbf{X})$
\item simulate $\mathbf{Z}' \sim p(\mathbf{Z}|\mathbf{S}',\mathbf{\theta},\mathbf{X})$
\item simulate $\mathbf{\theta}' \sim p(\mathbf{\theta}|\mathbf{Z}',\mathbf{S}',\mathbf{X})$
\end{enumerate}
\end{itemize}
Below we give the expressions of related conditional probability distributions.

\begin{itemize}

\item Sampling $\mathbf{Z} \sim p(\mathbf{Z}|\mathbf{X},\mathbf{S},\mathbf{\theta}) \propto 
p(\mathbf{X}|\mathbf{Z},\mathbf{\theta})p(\mathbf{Z})$:
\begin{eqnarray*}
p(\mathbf{X}|\mathbf{Z},\mathbf{\theta}) & = & \prod_r p(\mathbf{x}(r)|\mathbf{z}(r),\mathbf{\theta}) \\ 
& = & \prod_r \mathcal{N}(\mathbf{A}\mathbf{m}_{\mathbf{z}(r)},\mathbf{A}\Sigma_{\mathbf{z}(r)}\mathbf{A}^t+\mathbf{R}_\epsilon)
\end{eqnarray*}
where, $\mathbf{m}_{\mathbf{z}(r)} = [\mu_{1z_1(r)},\ldots,\mu_{Nz_N(r)}]^t$ and $\Sigma_{\mathbf{z}(r)} = diag[\sigma^2_{1z_1(r)},\ldots,\sigma^2_{Nz_N(r)}]$.

Notice $p(\mathbf{Z}) = \prod^N_{j=1} p(\mathbf{z}_j)$, and as mentioned earlier, $p(\mathbf{z}_j)$ takes the form of Potts MRF as (\ref{eqn:potts}). An inner Gibbs sampling is then used to simulate ${\mathbf{z}_j}$ with the likelihood $p(\mathbf{x}(r)|\mathbf{z}(r),\mathbf{A},\mathbf{\theta})$ marginalized over all configurations of $\{\mathbf{z}_{j'}(r),j' \neq j\}$.

\item Sampling $\mathbf{S} \sim p(\mathbf{S}|\mathbf{X},\mathbf{Z},\mathbf{\theta})$:
\begin{eqnarray*}
p(\mathbf{S}|\mathbf{X},\mathbf{Z},\mathbf{\theta}) & \propto & 
p(\mathbf{X}|\mathbf{S},\mathbf{A},\mathbf{R}_\epsilon)p(\mathbf{S}|\mathbf{Z},\mathbf{\theta}) \\
& = & \prod_r \mathcal{N}(\mathbf{m}^{apost}_s(r),\mathbf{R}^{apost}_s(r))
\end{eqnarray*}
$$ \left\{ \begin{array}{rcl}
\mathbf{R}^{apost}_s(r) & = & \left[ \mathbf{A}^t \mathbf{R}_\epsilon^{-1} \mathbf{A} + \Sigma_{\mathbf{z}(r)}^{-1} \right]^{-1} \\
\mathbf{m}^{apost}_s(r) & = & \mathbf{R}^{apost}_s(r) \left[ \mathbf{A}^t \mathbf{R}_\epsilon^{-1} \mathbf{x}(r) + \Sigma_{\mathbf{z}(r)}^{-1} \mathbf{m}_{\mathbf{z}(r)} \right]
\end{array} \right. $$

\item Sampling $\mathbf{R}_\epsilon$: 
$$p(\mathbf{R}_\epsilon|\mathbf{X},\mathbf{S},\mathbf{A}) \propto p(\mathbf{X}|\mathbf{S},\mathbf{A},\mathbf{R}_\epsilon)p(\mathbf{R}_\epsilon)$$
Considering we assign an inverse Wishart distribution to $p(\mathbf{R}_\epsilon)$, which is conjugate prior for the likelihood (\ref{eqn:datagaus}), $\mathbf{R}_\epsilon$ is \emph{a posteriori} sampled from:
$$ \left\{ \begin{array}{l}
\mathbf{R}_\epsilon^{-1} \sim \mathcal{W}_i(\alpha_\epsilon,\beta_\epsilon) \\
\alpha_\epsilon = \frac{1}{2}(|\mathcal{R}|-n), \beta_\epsilon = \frac{1}{2}|\mathcal{R}|(\mathbf{R}_{xx}-\mathbf{R}_{xs}\mathbf{R}_{ss}^{-1}\mathbf{R}_{xs}^t) \\
\end{array} \right. $$
where, the sample statistics 
$ \mathbf{R}_{xx} = \frac{1}{|\mathcal{R}|}\sum_r\mathbf{x}_r\mathbf{x}_r^t $, 
$ \mathbf{R}_{xs} = \frac{1}{|\mathcal{R}|}\sum_r\mathbf{x}_r\mathbf{s}_r^t $,
$ \mathbf{R}_{ss} = \frac{1}{|\mathcal{R}|}\sum_r\mathbf{s}_r\mathbf{s}_r^t $.

\item Sampling $\mathbf{A} \sim p(\mathbf{A}|\mathbf{X},\mathbf{S},\mathbf{R}_\epsilon)$:
$$p(\mathbf{A}|\mathbf{X},\mathbf{S},\mathbf{R}_\epsilon) \propto p(\mathbf{X}|\mathbf{S},\mathbf{A},\mathbf{R}_\epsilon)p(\mathbf{A})$$
Given uniform or Gaussian prior for $\mathbf{A}$, the posterior distribution of $\mathbf{A}$ is a Gaussian: 
$$ \left\{ \begin{array}{l}
\mathit{Vec}(\mathbf{A}) \sim \mathcal{N}(\mathbf{\mu}_A,\mathbf{R}_A) \\
\mathbf{\mu}_A = \mathit{Vec}(\mathbf{R}_{xs}\mathbf{R}_{ss}^{-1}), \mathbf{R}_A = \frac{1}{|\mathcal{R}|}\mathbf{R}_{ss}^{-1}\otimes\mathbf{R}_\epsilon \\
\end{array} \right. $$
where $\otimes$ is the Kronecker product and $\mathit{Vec(.)}$ represents the column-stacking operation. 

\item Sampling $(\mu_{jk},\sigma^2_{jk})$: \\
With $(Z,S)$ sampled in earlier steps and conjugate priors assigned, the means $\mu_{jk}$ 
and the variances $\sigma^2_{jk}$ can be sampled from respective posteriors as follows:
$$ \left\{ \begin{array}{rcl}
\mu_{jk}|\mathbf{s}_j,\mathbf{z}_j,\sigma^2_{jk} & \sim & \mathcal{N}(m_{jk},v_{jk}^2) \\
m_{jk} & = & v_{jk}^2 \left( \frac{\mu_{k0}}{\sigma_{k0}^2} + \frac{1}{\sigma^2_{jk}}\sum_{r\in\mathcal{R}^{(j)}_k}\mathbf{s}_j(r) \right) \\
v_{jk}^2 & = & \left( \frac{n^{(j)}_k}{\sigma^2_{jk}} + \frac{1}{\sigma_{k0}^2} \right)^{-1} \\
\end{array} \right. $$
and,
$$ \left\{ \begin{array}{rcl}
\sigma_{jk}^2|\mathbf{s}_j,\mathbf{z}_j,\mu_{jk} & \sim & \mathcal{IG}(\alpha_{jk},\beta_{jk}) \\
\alpha_{jk} & = & \alpha_{k0} + \frac{n^{(j)}_k}{2} \\
\beta_{jk} & = & \beta_{k0} + \frac{1}{2}\sum_{r\in\mathcal{R}^{(j)}_k}(\mathbf{s}_j(r)-\mu_{jk})^2 \\
\end{array} \right. $$
where, label region $\mathcal{R}^{(j)}_k = \{r:\mathbf{z}_j(r)=k\}$ and the region size $n^{(j)}_k = |\mathcal{R}^{(j)}_k|$.

\end{itemize}

\section{\uppercase{Simulation Results}}

\begin{figure}[!h]
  \centering
  \includegraphics[scale=0.35]{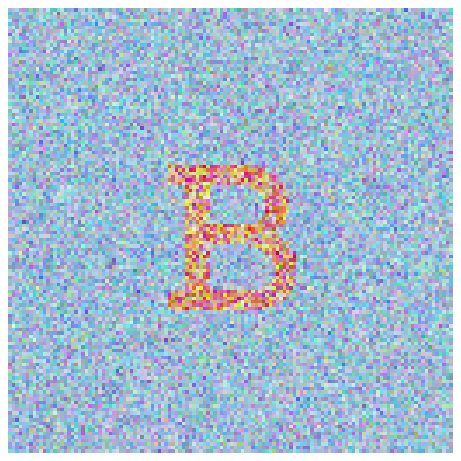} 
  \includegraphics[scale=0.35]{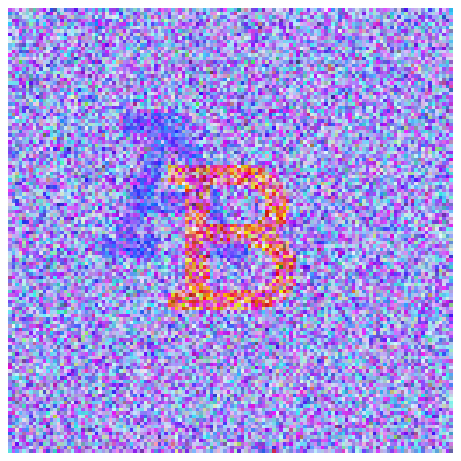} 
  \includegraphics[scale=0.35]{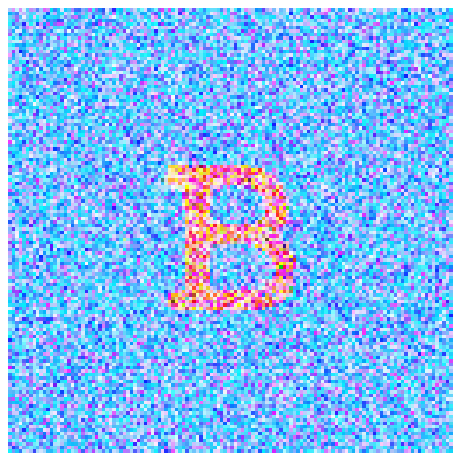} 
  \includegraphics[scale=0.35]{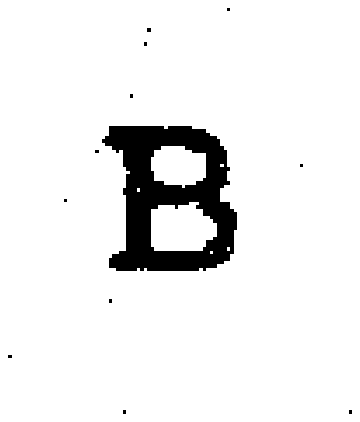} 
  \\
  \includegraphics[scale=0.35]{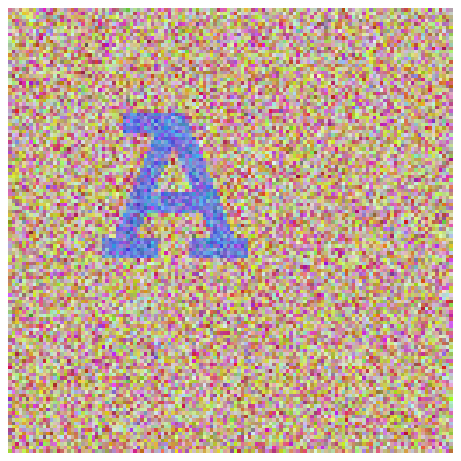} 
  \includegraphics[scale=0.35]{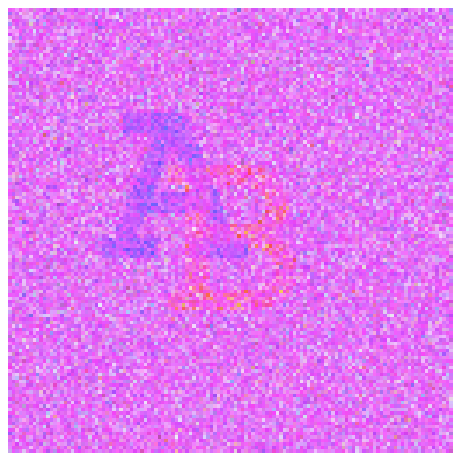} 
  \includegraphics[scale=0.35]{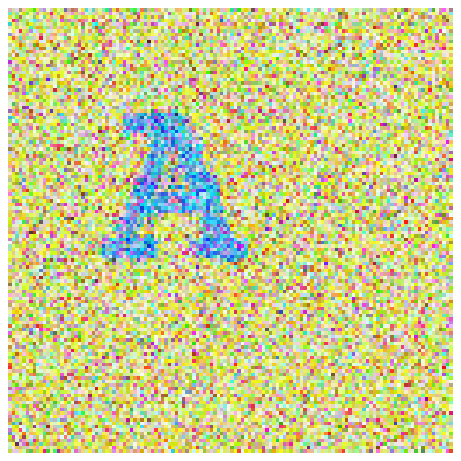} 
  \includegraphics[scale=0.35]{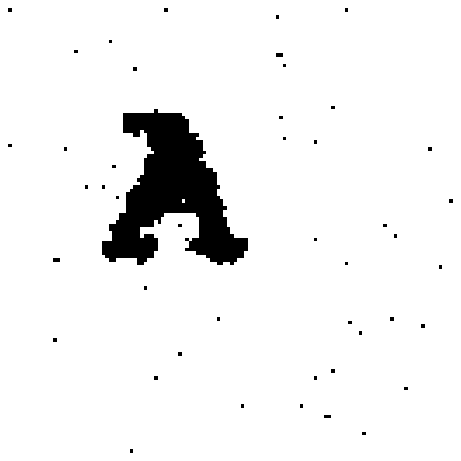} 
  \\
  (a) \hspace{1.0cm} (b) \hspace{1.0cm} (c) \hspace{1.0cm} (d)
  \caption{Separation of synthetic image mixtures: a) original sources; b) image mixtures; c) demixed sources; d) classification labels.}
  \label{fig:synsamp}
\end{figure}

\noindent For evaluating the performance of the proposed algorithm, we use both synthetic and real images in the test. The synthetic images were generated according to the model setting that each source is composed of pixels of two classes (text and background) and two source images are linearly mixed in every color channel independently to produce two observation images. This was done in three steps: 
\begin{enumerate}
\item Two binary ($K_{j=1,2}=2$) text image were scanned from real documents or created by graphic tools. They were used as the class labels $Z_{j=1,2}$ for each source;
\item With known means and variances for pixel value of each class, the source images were generated according to (\ref{eqn:srclab});
\item For each color channel, a random selected $\mathbf{A}_{2 \times 2}$ was used to mix the sources and finally white Gaussian noises $\mathbf{R}_\epsilon$ were added (SNR=20dB).
\end{enumerate}
Fig.\ref{fig:synsamp} shows the synthetic image mixtures, demixed sources and the label fields.

\begin{figure}[!h]
  \centering
  \includegraphics[scale=0.4]{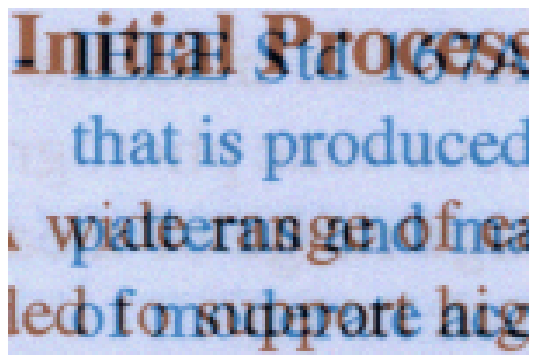} 
  \includegraphics[scale=0.4]{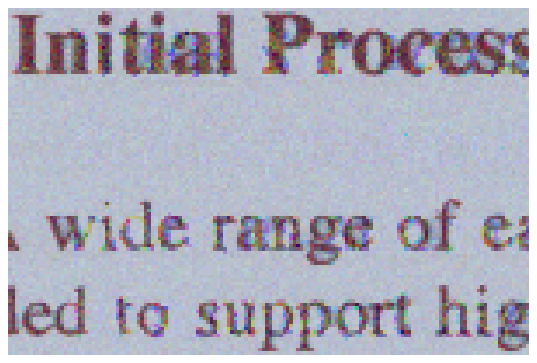} 
  \includegraphics[scale=0.4]{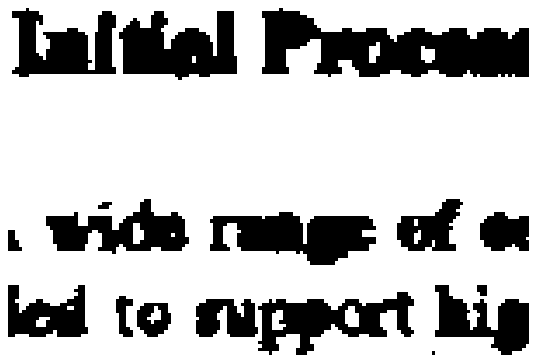} 
  \\
  \includegraphics[scale=0.4]{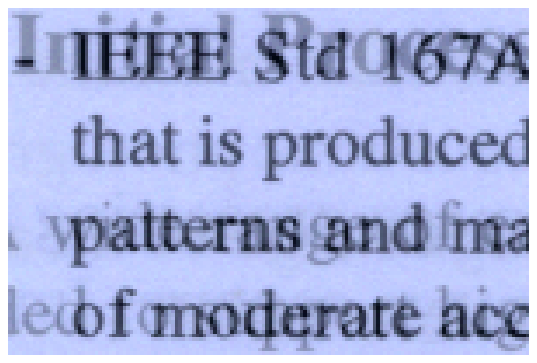} 
  \includegraphics[scale=0.4]{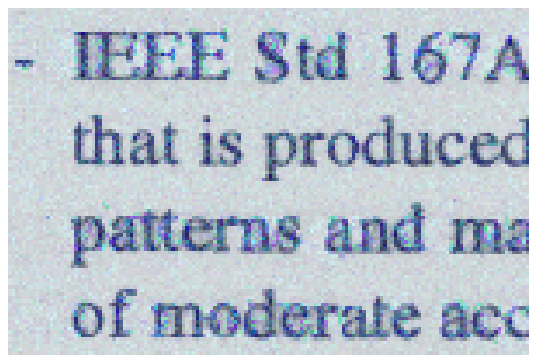} 
  \includegraphics[scale=0.4]{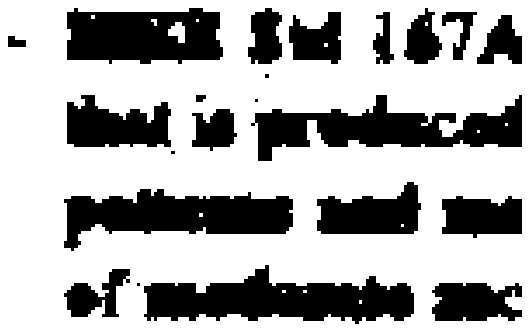} 
  \\
  (a) \hspace{1.5cm} (b) \hspace{1.5cm} (c)
  \caption{Separation of real show-through image mixtures: a) image mixtures; b) demixed sources; c) classification labels.}
  \label{fig:realsamp}
\end{figure}

The real image for test was scanned from a duplex printed paper, where show-through causes the superimposition of text. The separation result is shown in Fig.\ref{fig:realsamp}.

\begin{figure}[!h]
  \centering
  \includegraphics[scale=0.45]{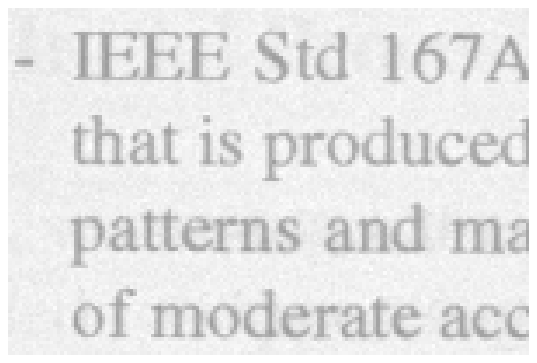} 
  \includegraphics[scale=0.45]{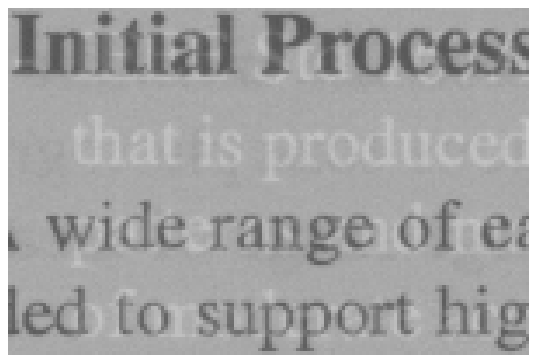} 
  \\
  (a) two ICs corresponding to the demixed sources
  \\
  \includegraphics[scale=0.33]{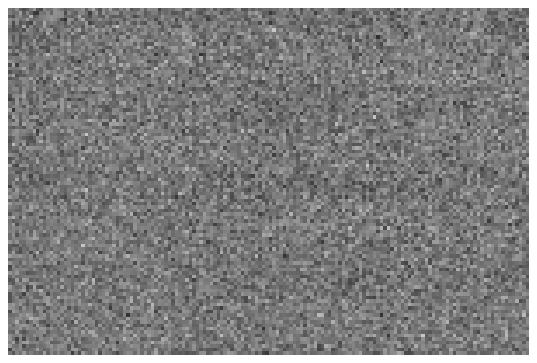} 
  \includegraphics[scale=0.33]{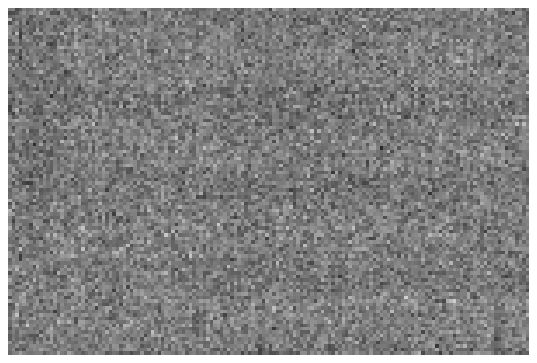} 
  \includegraphics[scale=0.33]{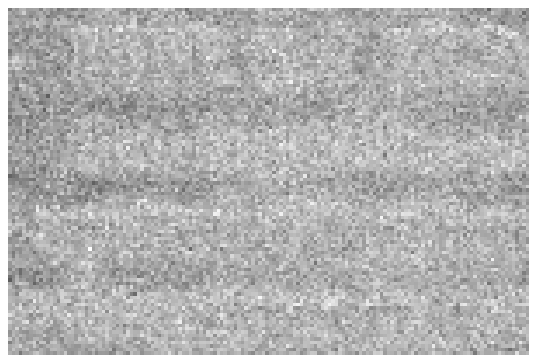} 
  \includegraphics[scale=0.33]{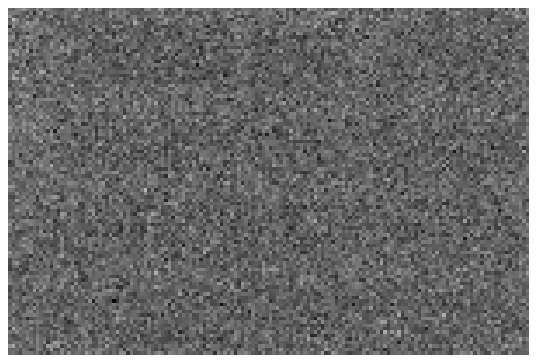} 
  \\
  (b) other ICs containing noise-like signals
  \caption{Separation results by ICA.}
  \label{fig:realsampica}
\end{figure}

For comparison, we also employed the FastICA algorithm \cite{Hyvarinen_TNN99} on the sample images with typical parameter set. The results on the show-through examples of Fig.\ref{fig:realsamp} are shown in Fig.\ref{fig:realsampica}. All three channels of the two observed mixtures were used as inputs simultaneously to the ICA algorithm. The two demixed sources can be found in two of six independent components (IC) outputed, while the other four output ICs usually contain unintended noise-like signals, which, along with the permutability property of the ICA algorithm, bring difficulties to reconstructing color representation of the sources. On the other hand, when less color channels are exploited in demixing, we noticed that the separation result does not necessarily degrade or improve, owing to the possible presence of cross-channel correlations.

The MCMC computation involved in the proposed Bayesian separation method is time-consuming. For the example image of 300x240 pixels in Fig.\ref{fig:realsamp}, which is small relative to ordinary document sizes and resolutions, the typical computation time of the experimental implementation can come up to hours without specific optimizations. However, various computing alternatives such as Mean Field and variational approximation can be exploited to achieve higher efficiency.

\section{\uppercase{Conclusion}}
\label{sec:conclusion}

\noindent We proposed a Bayesian approach for separating noisy linear mixture of document images. For source images, we considered a hierarchical model with the hidden label variable $\mathbf{z}$ representing the common classification of objects among multiple color channels, and a Potts-Markov prior was employed for the class labels imposing local regularity constraints. We showed how Bayesian estimation of all unknowns of interest can be computed by MCMC sampling from their posterior distributions given the observation. We then illustrated the feasibility of the proposed algorithm on joint separation and segmentation by tests on sample images.


\renewcommand{\baselinestretch}{0.98}
\bibliographystyle{apalike}
{\small
\bibliography{bssref}}
\renewcommand{\baselinestretch}{1}

\end{document}